\documentclass[aps,pra,twocolumn]{revtex4}
\usepackage[utf8]{inputenc}
\usepackage[normalem]{ulem}
\usepackage{xcolor}
\usepackage{graphicx}
\usepackage{bm}
\usepackage{amsmath}
\usepackage[colorlinks=true,allcolors=blue]{hyperref}

\providecommand{\ket}[1]{\lvert #1 \rangle}

\providecommand{\be}{\begin{equation}}
\providecommand{\ee}{\end{equation}}
\providecommand{\ba}{\begin{eqnarray}}
\providecommand{\ea}{\end{eqnarray}}

\usepackage[makeroom]{cancel}

\begin{document}

\title{Resonator-assisted quantum transduction between superconducting qubits and trapped atomic systems via Rydberg levels}

\author{Fernando L. Semi\~ao}
\affiliation{Centro de Ci\^encias Naturais e Humanas, Universidade Federal do ABC - UFABC, Santo Andr\'e, Brazil}
\author{Matthias Keller}
\affiliation{Department of Physics and Astronomy, University of Sussex, Brighton BN1 9RH, United Kingdom}

\begin{abstract}
Using a shared microwave resonator, we propose a transduction scheme between superconducting qubits and qubit states encoded in the low-lying internal levels of trapped atomic systems. The approach employs atomic Rydberg levels together with laser pulses that connect them to the low-lying qubit states. We explore two coupling protocols: one based on resonant interactions between the subsystems, and another operating in the dispersive regime, where the resonator is far detuned from the transition frequencies of the matter qubits. These protocols enable the transfer of a general qubit state from the superconducting qubit to the atomic qubit. Transfer fidelity is evaluated across different coupling regimes and under various sources of dissipation and decoherence, including resonator decay, Rydberg state decay, and both relaxation and pure dephasing of the superconducting qubit, providing a detailed assessment of the performance of the proposed scheme.
\end{abstract}

\maketitle

\section{Introduction}
The proposed transducers between microwave and optical systems cover a variety of approaches, including atomic systems \cite{PhysRevLett.102.083602,PhysRevLett.103.043603,PhysRevA.85.020302}, ions in solid-state crystals \cite{PhysRevLett.113.063603,PhysRevLett.113.203601,PhysRevA.92.062313}, optomechanical systems \cite{PhysRevLett.105.220501,Andrews2014-bo,Bagci2014-it,PhysRevLett.109.130503,Safavi-Naeini_2011,PhysRevA.82.053806,PhysRevLett.108.153603,PhysRevLett.108.153604,Hill2012-y}, electro-optical systems \cite{PhysRevA.81.063837,PhysRevA.96.043808}, and several other designs \cite{PhysRevB.93.174427,PhysRevA.99.063830,PhysRevB.101.214414,PhysRevLett.118.140501,PhysRevA.100.053843,PhysRevB.96.165312, Hensinger, molecular,milburn}. Current interest in this problem stems from the potential to harness and combine the individual strengths of dissimilar systems. The prospect of enabling state transduction and entanglement between superconducting qubits and trapped atomic systems is particularly compelling, as it brings together two of the most successful quantum computing platforms to date, offering both high scalability and integration potential. Moreover, these platforms are highly complementary: superconducting qubits allow for fast gate operations and precise electronic control, while atoms and ions exhibit excellent coherence times and facilitate efficient state transfer to photonic qubits. In particular, the ability to couple atomic states to photons with high efficiency enables the transmission of quantum information across quantum networks, paving the way for distributed quantum computing \cite{Krutyanskiy, Keller}.

A promising approach to integrating trapped atomic systems with superconducting qubits is to exploit atomic Rydberg levels, which offer transition frequencies in the microwave range, long lifetimes, and large dipole moments \cite{haroche}. These properties have enabled, for example, the control of entanglement between neutral atoms \cite{wilk}, as well as the implementation of quantum logic gates between individually addressed atomic qubits \cite{Saffman}. In the case of Rydberg ions, there have been demonstrations of coherent excitation to Rydberg levels \cite{kaler}, state manipulation \cite{markusprl}, and entangling gates based on the strong dipole–dipole interaction between two trapped Rydberg ions \cite{markus}.

Here, we propose a transducer to effectively link microwave superconducting qubits to the low-lying levels of a trapped ion or atom. What sets our proposal apart from previous ones \cite{PhysRevLett.102.083602,PhysRevLett.103.043603,PhysRevA.85.020302,PhysRevLett.113.063603,PhysRevLett.113.203601,PhysRevA.92.062313,PhysRevLett.105.220501,Andrews2014-bo,Bagci2014-it,PhysRevLett.109.130503,Safavi-Naeini_2011,PhysRevA.82.053806,PhysRevLett.108.153603,PhysRevLett.108.153604,Hill2012-y,PhysRevA.81.063837,PhysRevA.96.043808,PhysRevB.93.174427,PhysRevA.99.063830,PhysRevB.101.214414,PhysRevLett.118.140501,PhysRevA.100.053843,PhysRevB.96.165312, Hensinger, molecular,milburn} is the use of a microwave resonator as a data bus \cite{haroche,sillanpaa,blais}, combined with laser pulses to couple Rydberg states to low-lying levels \cite{kaler,markusprl}. By merging these elements, we demonstrate the potential of this setup to transfer a general qubit state, initially encoded in the superconducting qubit, to the optically active levels of the trapped atom, which serve as the atomic qubit.

The microwave resonator can be implemented, for example, as a coplanar waveguide (CPW) \cite{sillanpaa}. Experimentally, the interaction between the CPW resonators and Rydberg levels was recently demonstrated \cite{Kaiser}. At the same time, coupling superconducting qubits to CPW cavities has become routine, with various control techniques available, such as the use of Purcell filters \cite{blais}. A key element of our proposal is the application of laser pulses to de-excite the atom from the populated Rydberg levels to low-lying qubit levels. For neutral atoms, coupling between low-lying levels and Rydberg states has a long history \cite{haroche}, and significant advances have been made in recent years for trapped ions \cite{markusprl,markus,kaler}. In our scheme, we use relatively intense deexcitation pulses to effectively switch off the resonator interaction by transferring the population out of the Rydberg levels. 

This paper is organized as follows. In Sect. \ref{set}, we introduce the composite system, which consists of a superconducting qubit, a microwave resonator, and a trapped atom. We present the system Hamiltonian and the figure of merit used to assess the quality of transduction, namely the quantum fidelity \cite{NC}. In Sect. \ref{irc}, we describe our first transduction protocol, which employs a resonator simultaneously resonant with the superconducting qubit and the atomic Rydberg levels. In Sect. \ref{idc}, we introduce a second transduction protocol, in which the resonator is detuned from the qubit transition frequencies of the matter qubits to induce a dispersive interaction between them. In Sect. \ref{noise}, we account for dissipation and decoherence and discuss their impact on the performance of the protocols. Finally, we present our conclusions in Sect. \ref{sumup}.
\section{The System}\label{set}
\begin{figure}[h!]
	\includegraphics[width=\linewidth]{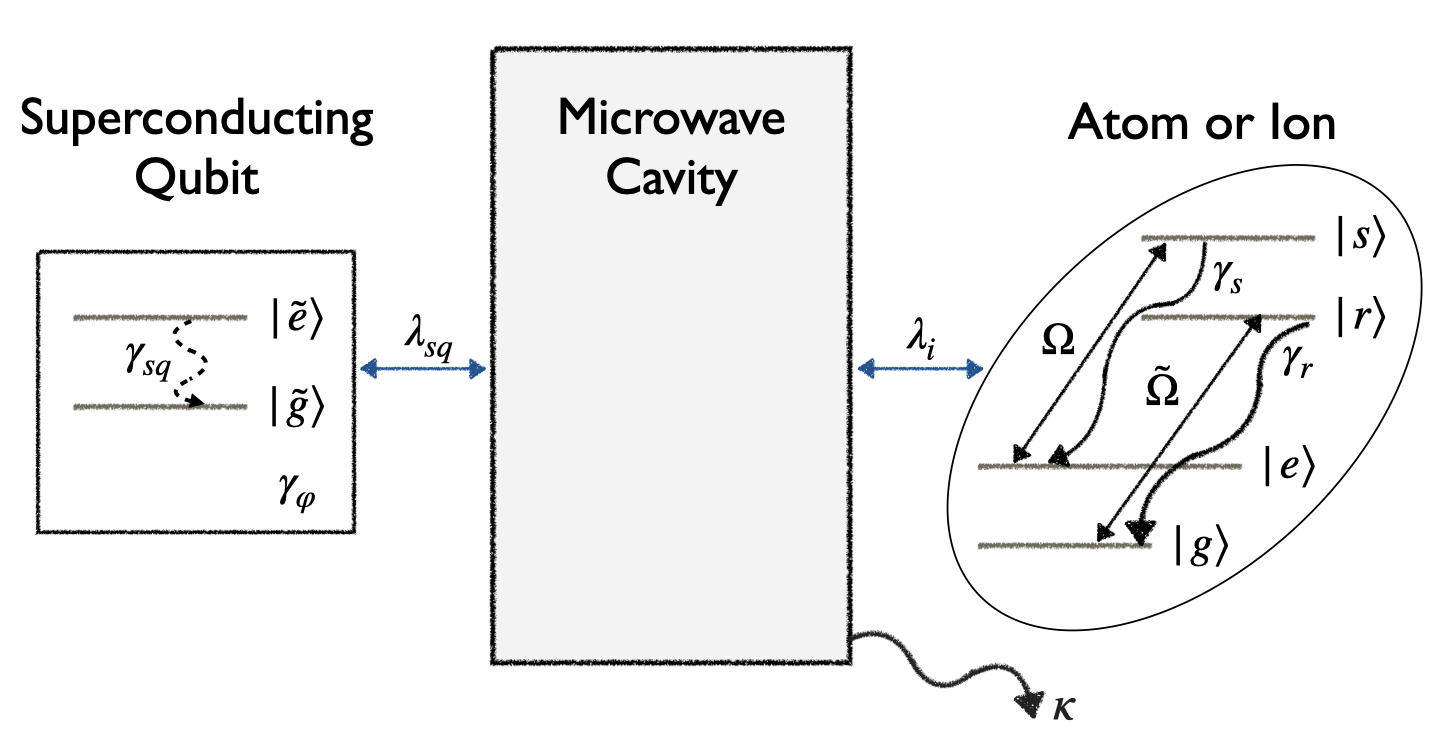}
	\caption{On the left, the superconducting qubit is represented by states $|\tilde g\rangle$ and $|\tilde e\rangle$, interacting with the microwave resonator described by the coupling constant $\lambda_{sq}$. In the center, the resonator mediates the interaction between the superconducting qubit and the atomic system. On the right, the atomic levels include the low-lying states $|g\rangle$ and $|e\rangle$, and the Rydberg states $|r\rangle$ and $|s\rangle$, connected by laser-driven transitions. The laser interaction is characterized by the Rabi frequencies $\Omega$ and $\tilde\Omega$. The Rydberg states couple to the resonator with strength $\lambda_i$. The model still includes: Rydberg state decay $\gamma_{s,r}$, resonator decay $\kappa$, as well as relaxation $\gamma_{sq}$ and dephasing $\gamma_\varphi$ of the superconducting qubit.} \label{setup}
\end{figure}

In this work, we explore the integration of three quantum systems, as illustrated in Fig.\ref{setup}. The superconducting qubit is modeled as a two-level system (left side of Fig.\ref{setup}), with an excited state $|\tilde e\rangle$ and a ground state $|\tilde g\rangle$, and is coupled to a resonator mode with coupling strength $\lambda_{sq}$. The atomic particle is modeled as a four-level system (right side of Fig.\ref{setup}), featuring two Rydberg states, $|r\rangle$ and $|s\rangle$, which couple to the same resonator mode with coupling strength $\lambda_i$, and two low-lying states, $|g\rangle$ and $|e\rangle$, which form the target qubit. Transitions between Rydberg and low-lying states are driven by controllable laser fields with Rabi frequencies $\Omega$ and $\tilde\Omega$ (see, e.g., \cite{markusprl,markus,Lahaye}). In Sect.\ref{noise}, the model accounts for spontaneous decay of the Rydberg states (rates $\gamma_{s,r}$), resonator decay (rate $\kappa$), and both relaxation ($\gamma_{sq}$) and dephasing ($\gamma_\varphi$) of the superconducting qubit.

We consider only the two spontaneous decay channels for the atomic system shown in  Fig.~\ref{setup}. By choosing the low-lying and Rydberg state parities correctly, single-photon spontaneous emissions from $|s\rangle$ to $|g\rangle$ and $|r\rangle$ to $|e\rangle$ can be suppressed. With multiphoton decays usually significantly weaker, we neglect these decay channels.

As seen in Fig. \ref{setup}, the atomic system interacts with the superconducting qubit only indirectly, through the bosonic link provided by the microwave resonator. The system Hamiltonian, when the lasers are turned off, consists of two simultaneous Jaynes-Cummings-like interactions, which can be expressed as follows (with $\hbar = 1$).
\begin{eqnarray}\label{H1}
H&=& E_g|g\rangle\langle g| +E_e|e\rangle\langle e|+E_r|r\rangle\langle r|+E_s|s\rangle\langle s| +\omega a^\dag a \nonumber \\
&&+\lambda_i(a^\dag|r\rangle\langle s|+a|s\rangle\langle r|)+E_{\tilde g}|{\tilde g}\rangle\langle {\tilde g}| +E_{\tilde e}|{\tilde e}\rangle\langle {\tilde e}|\nonumber\\
&&+\lambda_{sq}(a^\dag|\tilde g\rangle\langle \tilde e|+a|\tilde e\rangle\langle \tilde g|),
\end{eqnarray}
where $E_g < E_e < E_r < E_s$ are the electronic level energies of the atomic system and $E_{\tilde g} < E_{\tilde e}$ are the level energies of the superconducting qubit. The operators $a^\dag$ and $a$ represent the creation and annihilation operators for the resonator excitation and $\omega$ is the angular frequency of the resonator mode.

In the following, we will not focus on the specific values of the level energies or the resonator frequency, since only the transition frequency mismatch is relevant. However, coupling constants play a crucial role in determining the time scale of the protocol.  Typical coupling strengths between superconducting qubits and CPW resonators, $\lambda_{sq}$, range from tens to hundreds of MHz. For example, in \cite{kappa2}, the coupling constant between the superconducting qubit and the microwave resonator is $\lambda_{sq}/2\pi = 61$ MHz. For the dipole interaction between Rydberg levels and a microwave resonator, the coupling strengths are somewhat lower. For example, in \cite{Kaiser}, a coupling constant of $\lambda_i/2\pi = 8$ MHz was measured in a system consisting of an ultra-cold cloud of $^{87}$Rb atoms and a CPW microwave resonator.

Our goal with this system is to present feasible methods for transferring a general qubit state from a superconducting qubit, which operates in the microwave domain, to two low-lying electronic levels of a trapped atomic system. Specifically, after preparing the superconducting qubit in the state
\begin{eqnarray}\label{sqstate}
|sq\rangle = \cos(\theta/2)|\tilde e\rangle + e^{i\phi} \sin(\theta/2)|\tilde g\rangle,
\end{eqnarray}
and initializing both the resonator and the atom in appropriate initial states, we calculate the evolution of the system's density operator $\rho(t)$, trace out the resonator and the superconducting qubit, and obtain the atomic state $\rho_{\rm{at}}(t)$. This is a $4 \times 4$ density matrix defined in the space spanned by $\{|g\rangle, |e\rangle, |r\rangle, |s\rangle\}$, which is then compared to the target state
\begin{eqnarray} \label{tg}
|\text{target}\rangle = \cos(\theta/2) |e\rangle + e^{i\phi} \sin(\theta/2) |g\rangle + 0 |r\rangle + 0 |s\rangle.
\end{eqnarray}
To quantify how close the evolved state is to the target, we use the fidelity of the quantum state defined as \cite{NC}.
\begin{eqnarray}\label{F}
F(t) = \sqrt{\langle \text{target} |\rho_{\rm{at}}(t)|\text{target}\rangle}.
\end{eqnarray}

\section{Ideal Resonant Case} \label{irc}
First we consider the resonant protocol, in which the matter qubits and the microwave resonator are in resonance. This can be achieved, e.g. by adjusting the resonator length and Stark-shifting the Rydberg states. In this case, the Hamiltonian (\ref{H1}) in the interaction picture becomes time-independent and is given by
\begin{eqnarray}\label{H2}
H_R=\lambda_i(a^\dag|r\rangle\langle s|+a|s\rangle\langle r|)+\lambda_{sq}(a^\dag|\tilde g\rangle\langle \tilde e|+a|\tilde e\rangle\langle \tilde g|).
\end{eqnarray}  

Our transduction protocol consists of two steps. First, a general superposition in the superconducting qubit is transferred to the Rydberg states $\{|r\rangle, |s\rangle\}$, using the resonator as a quantum bus. Then, without removing the resonator interaction, laser pulses are applied to map the Rydberg quantum state onto the low-lying states, $|g\rangle$ and $|e\rangle$, which act as the target qubit states.

The protocol begins with the trapped atomic particle initially prepared in the Rydberg state $|r\rangle$, the resonator in the vacuum state $|0\rangle_c$, and the superconducting qubit in the general superposition state $|sq\rangle$, as given by Eq. (\ref{sqstate}). The initial state of this system can then be written as
\begin{eqnarray}\label{is}
|\psi(0)\rangle &=&|r\rangle\otimes|0\rangle_c\otimes\left[\cos(\theta/2) |\tilde e\rangle+e^{i\phi}\sin(\theta/2)|\tilde g\rangle \right]\nonumber\\
&=& \cos(\theta/2)|r,0,\tilde e\rangle+e^{i\phi}\sin(\theta/2)|r,0,\tilde g\rangle.
\end{eqnarray}
Under the Hamiltonian in Eq. (\ref{H2}), and without considering any decoherence yet, the initial state in Eq. (\ref{is}) evolves unitarily to
\begin{eqnarray}\label{psit}
|\psi(t)\rangle &=& c_e(t)|r,0,\tilde e\rangle+c_g(t)|r,0,\tilde g\rangle + \alpha_1(t)|r,1,\tilde g\rangle \nonumber\\   &&+ \alpha_0(t)|s,0,\tilde g\rangle
\end{eqnarray}
with
\begin{eqnarray}
c_e(t)&=&\cos(\theta/2)\,\frac{(\lambda_{sq}^2\cos\tilde\lambda t+\lambda_i^2)}{\tilde\lambda^2},\nonumber\\
c_g(t)&=&e^{i\phi}\sin(\theta/2)\nonumber,\\
\alpha_1(t)&=&-i\cos(\theta/2)\,\frac{\lambda_{sq}\sin\tilde\lambda t}{\tilde\lambda},\nonumber\\
\alpha_0(t)&=&-\cos(\theta/2)\,\frac{2\lambda_i\lambda_{sq}}{\tilde\lambda^2}\sin^2\left(\frac{\tilde\lambda t}{2}\right),
\label{eq_StateVector}
\end{eqnarray}
where $\tilde\lambda=\sqrt{\lambda_i^2+\lambda_{sq}^2}$. 

We aim for the original parameters $\theta$ and $\phi$ to be transferred to a superposition of the Rydberg states $|r\rangle$ and $|s\rangle$, with the states of the superconducting qubit and resonator factored out. This is achieved when both $c_e(t)$ and $\alpha_1(t)$ vanish simultaneously. In this case, no population has leaked from $|\tilde e\rangle$ to $|r\rangle$, no excitation remains in the resonator, and there is no residual entanglement with the superconducting qubit. By imposing  $c_e(\tau_R) = 0$ and $\alpha_1(\tau_R) = 0$ on (\ref{eq_StateVector}), or equivalently $|c_g(\tau_R)|^2 + |\alpha_0(\tau_R)|^2 = 1$, the interaction time is obtained:
\begin{eqnarray}
\tau_R=\frac{2}{\tilde\lambda}\arcsin{\left(\sqrt[4]{\frac{\tilde\lambda^4}{4\lambda_i^2\lambda_{sq}^2}}\right)}.
\end{eqnarray}
Choosing $a_1 = b_2 = \sqrt{\lambda_i^2}$ and $a_2 = b_1 = \sqrt{\lambda_{sq}^2}$ in the Cauchy-Schwartz inequality $(a_1b_1+a_2b_2)^2\leq(a_1^2+a_2^2)(b_1^2+b_2^2)$, valid for any real $a_1,a_2,b_1$ and $b_2$, 
results in $4\lambda_i^2\lambda_{sq}^2\leq\tilde\lambda^4$, with equality achieved if and only if $\lambda_i = \lambda_{sq}$. 
Since the range of the arcsine function is $[-1, 1]$, $\tau_R$ is a real number if and only if $\lambda_i = \lambda_{sq} = \lambda$. Therefore, $\tau_R = \frac{\pi}{\sqrt{2}\lambda}$. The solutions are periodic; however, longer times are of limited interest because of the detrimental effects of decoherence, as discussed later. 

After allowing the system to evolve for a time $\tau_R$, one then finds  
\begin{eqnarray}\label{psi2}
|\psi\rangle_{\tau_R} = \left[e^{i\phi}\sin(\theta/2)|r\rangle-\cos(\theta/2)|s\rangle\right]  \otimes |0\rangle \otimes|\tilde g\rangle.
\end{eqnarray}
The state in Eq.~(\ref{psi2}) serves as the initial state for the second step of the protocol, in which classical laser pulses transfer the population and coherences to the lower levels $|e\rangle$ and $|g\rangle$. This is accomplished using two resonant laser fields: one driving the $|s\rangle \leftrightarrow |e\rangle$ transition with Rabi frequency $\Omega$, and the other driving $|r\rangle \leftrightarrow |g\rangle$ with Rabi frequency $\tilde\Omega$. These fields implement rotations $|s\rangle \to -i|e\rangle$ and $|r\rangle \to i|g\rangle$, respectively, when applied for a duration $T_R$ such that $\Omega T_R = \pi/2$ and $\tilde\Omega T_R = 3\pi/2$, which implies $\tilde\Omega = 3\Omega$. The final state after this step is
\begin{eqnarray}\label{f1}
|\psi\rangle_R=i\left[\cos(\theta/2)|e\rangle+e^{i\phi}\sin(\theta/2)|g\rangle\right]\otimes|0\rangle\otimes|\tilde g\rangle.
\end{eqnarray}
Apart from an irrelevant global phase, this is a perfect transduction, i.e. $|\psi\rangle_R = |\text{target}\rangle\otimes|0\rangle\otimes|\tilde g\rangle$. 

The state in Eq.~(\ref{f1}) is obtained under the idealized assumption that the interactions of both the atom and the superconducting qubit with the resonator are completely turned off during the application of the laser pulses. However, in practice, fully suppressing these interactions is often unfeasible. It is therefore crucial to assess how residual coupling to the resonator affects the state transfer. To this end, we consider the following time-dependent Hamiltonian
\begin{eqnarray} \label{ht}
H(t) &=& H_R + u(t-\tau_{R})\Omega (|e\rangle \langle s| + |s\rangle \langle e|) \nonumber \\ && + u(t-\tau_{R})\tilde\Omega (|r\rangle \langle g| + |g\rangle \langle r|), 
\end{eqnarray}
where $u(x)$ is the Heaviside function, $H_R$ is the Hamiltonian in Eq.(\ref{H2}), and $\tau_R$ is the time after which the lasers are turned on. For the simulations below, we use this time-dependent Hamiltonian in the von Neumann equation $\dot{\rho} = -i[H(t), \rho]$ and solve it to $0\leq t \leq \tau_R + T_R$.

\begin{figure}[t!]
    \includegraphics[width=\linewidth]{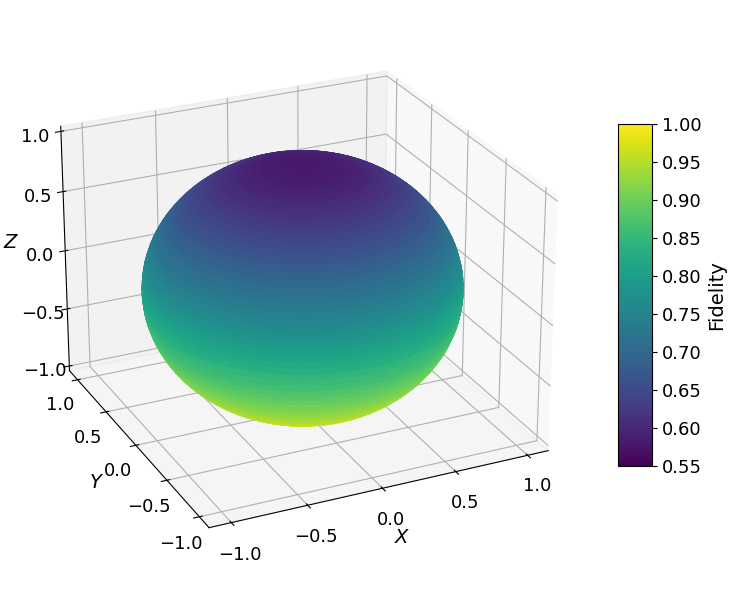}
    \includegraphics[width=\linewidth]{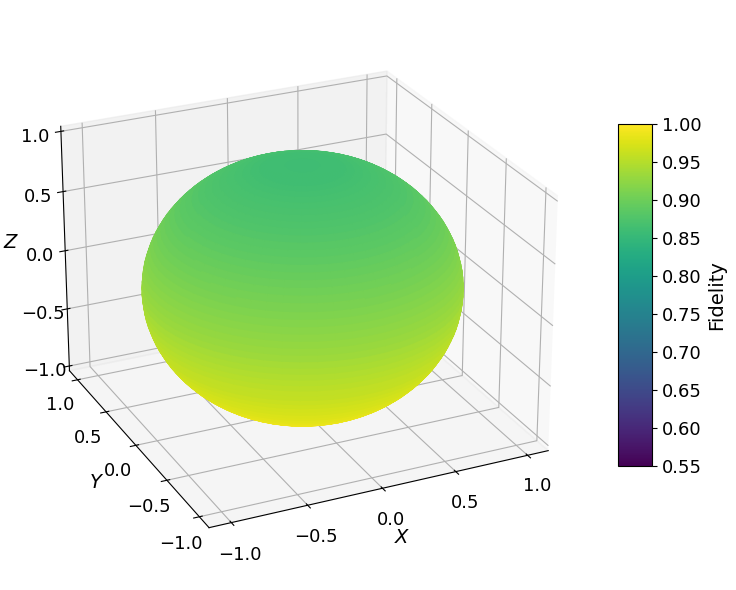}
     \caption{Bloch sphere visualization of the fidelity of the resonant protocol at $t = \tau_R + T_R$ for $\lambda = 2\pi \times 8$ MHz \cite{Kaiser}, with $\Omega / 2\pi = 4.8$ MHz \cite{Lahaye} ($\Omega / \lambda = 0.6$) in the top panel and $8$ MHz ($\Omega / \lambda = 1$) in the bottom panel.}	\label{bloch_r}
\end{figure}

In Fig.~\ref{bloch_r}, we present the Bloch sphere representation of the fidelity, which visually illustrates the performance of the protocol for all possible input states. We focus on the role of the ratio $\Omega / \lambda$, as it determines how quickly the laser pulses transfer population from the Rydberg states $\ket{s}$ and $\ket{r}$ to the low-lying states $\ket{e}$ and $\ket{g}$, while the rest of the system continues to evolve. If the population transfer induced by the laser pulses is too slow, the resonator tends to repopulate, which negatively affects the fidelity. For simulations, we assume $\lambda_i / 2\pi = 8$ MHz, a realistic reference value~\cite{Kaiser}, as discussed earlier. In the upper panel, we set $\Omega / 2\pi = 4.8$ MHz~\cite{Lahaye}, corresponding to $\Omega / \lambda = 0.6$, while in the lower panel, we increase the Rabi frequency to $\Omega / 2\pi = 8$ MHz, resulting in $\Omega / \lambda = 1$.

For $\Omega / \lambda = 0.6$ the resonator interaction strongly influences the dynamics, resulting in a minimum fidelity of $F_{\text{min}} = 0.58$. This is not unexpected, since the atom-resonator interaction is faster than the state transfer into the low-lying atomic states. Surprisingly, the protocol performs significantly better for only moderate increases in $\Omega$, as shown in the lower panel of Fig.~\ref{bloch_r}. By increasing $\Omega / \lambda$ from $0.6$ to $1$, the minimum fidelity increases significantly to $F_{\text{min}} = 0.86$. Extrapolating to $\Omega / \lambda = 3$ yields $F_{\text{min}} = 0.986$, and $\Omega / \lambda = 4$ would result in near-perfect transduction with $F_{\text{min}} = 0.995$, making our protocol robust against resonator dynamics during the application of laser pulses. Coherent state transfer between low-lying and Rydberg states using pulsed lasers has been demonstrated with Rabi frequencies in the GHz range \cite{Chew, Huber}. Although the resulting transfer fidelities remain limited, these experiments suggest that the increased values of $\Omega$ assumed in this work are within experimental reach. 

\begin{figure}[]
\includegraphics[width=\linewidth]{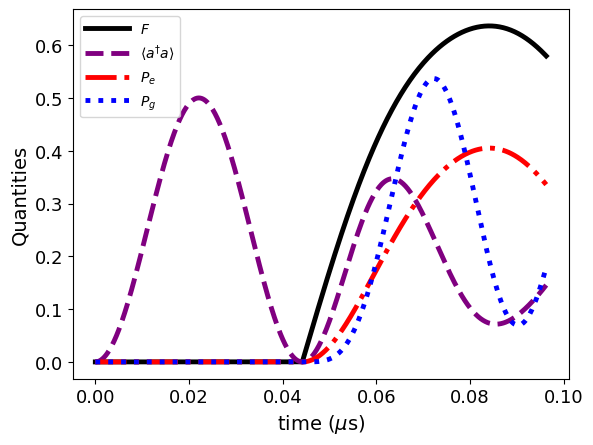}
\includegraphics[width=\linewidth]{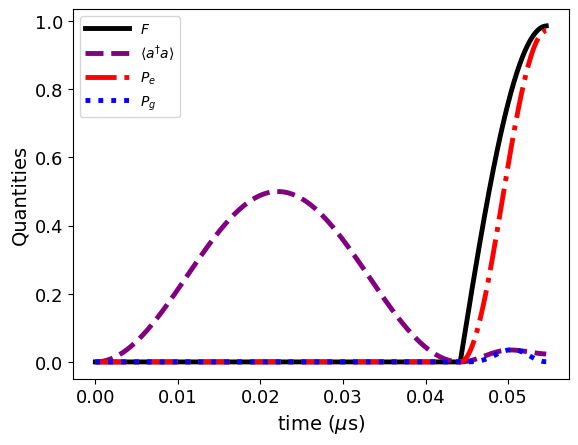}
\caption{Dynamics of key quantities for the resonant transduction protocol over the time interval $0 \leq t \leq \tau_R + T_R$: fidelity (solid black), mean resonator photon number (dashed purple), population in $|e\rangle$ (dash-dotted red), and population in $|g\rangle$ (dotted blue). The initial state is given by Eq. (\ref{is}) with $\theta = 0$. We set $\lambda = 2\pi \times 8$ MHz. Top panel: $\Omega / 2\pi = 4.8$ MHz ($\Omega / \lambda = 0.6$). Bottom panel: $\Omega / 2\pi = 24$ MHz ($\Omega / \lambda = 3$).
} \label{ideal_e}
\end{figure}

To gain a better understanding of the dynamics of the system, we analyze in more detail the evolution of the fidelity and state populations for two representative initial conditions given by Eq.~(\ref{is}): $\theta = 0$ (no superposition) and $\theta = \pi/2$, $\phi = 0$ (coherent superposition). The first case concerns the transfer of populations through the system, while the second focuses on the transfer of populations and coherences. Unless explicitly stated otherwise, we use $\lambda/2\pi = 8$ MHz throughout the simulations. In Fig.~\ref{ideal_e}, we consider the first case ($\theta = 0$), with the top panel corresponding to $\Omega / \lambda = 0.6$ and the bottom panel corresponding to an increased Rabi frequency ($\Omega / \lambda = 3$). During the initial part of the protocol, the populations $P_g$ and $P_e$ of the levels $|g\rangle$ and $|e\rangle$, respectively, remain zero. This holds until around $0.044~\mu$s, which marks the moment when the transfer lasers are turned on. After this, $P_g$ should ideally remain zero, while $P_e$ should evolve to one at the end of the protocol. It is clear that $\Omega / \lambda = 0.6$ is insufficient to achieve this, resulting in relatively low fidelity. On the other hand, $\Omega / \lambda = 3$ performs much better, reaching $P_e = 0.972$ and $F = 0.986$. This difference in performance can be understood by examining the dynamics of the resonator mode, as seen in Fig.~\ref{ideal_e}. Ideally, the mean number of resonator photons $\langle a^\dag a \rangle$ should remain zero after the lasers are turned on, which is clearly not the case when $\Omega \lesssim \lambda$.

\begin{figure}[]
    \includegraphics[width=\linewidth]{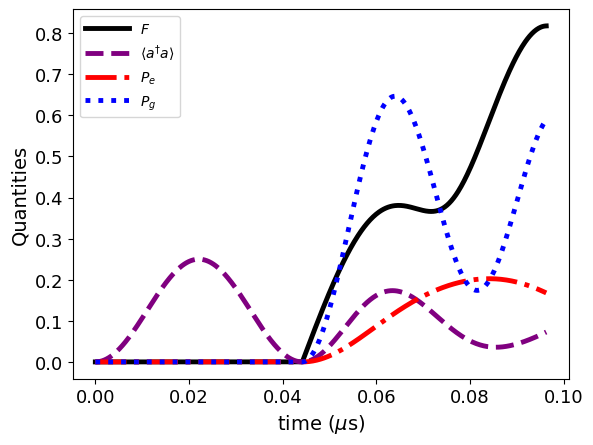}
     \includegraphics[width=\linewidth]{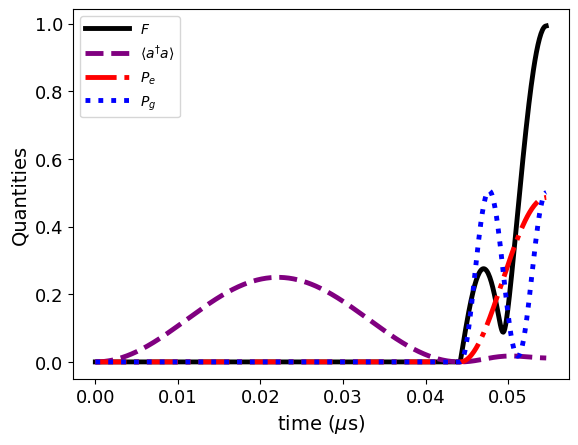}
     \caption{Dynamics of key quantities for the resonant transduction protocol over the time interval $0 \leq t \leq \tau_R + T_R$: fidelity (solid black), mean resonator photon number (dashed purple), population in $|e\rangle$ (dash-dotted red), and population in $|g\rangle$ (dotted blue). The initial state is given by Eq. (\ref{is}) with $\theta = \pi/2$, $\phi = 0$. We set $\lambda = 2\pi \times 8$ MHz. Top panel: $\Omega / 2\pi = 4.8$ MHz ($\Omega / \lambda = 0.6$). Bottom panel: $\Omega / 2\pi = 24$ MHz ($\Omega / \lambda = 3$).}
	\label{ideal_eg}
\end{figure}

In Fig. \ref{ideal_eg}, we consider the second representative initial state in Eq.(\ref{is}), corresponding to the coherent superposition $\theta = \pi/2$, $\phi = 0$. As before, we set $\lambda/2\pi = 8$ MHz for both panels, with $\Omega / \lambda = 0.6$ in the top panel and $\Omega / \lambda = 3$ in the bottom panel. After the lasers are turned on, the populations $P_e$ and $P_g$ should each reach $0.5$ at the end of the protocol, reflecting the equal-weight superposition. For $\Omega / \lambda = 0.6$, this is not achieved, and the protocol results in $P_e = 0.168$ and $F = 0.818$. However, for $\Omega / \lambda = 3$, we obtain $P_e = 0.468$ and $F = 0.993$. This improvement is again correlated with the suppression of the resonator dynamics when the Rabi frequency $\Omega$ exceeds the resonator coupling $\lambda$. In the limit $\Omega \gg \lambda$, the protocol produces perfect transduction regardless of the values of $\theta$ and $\phi$ in Eq.~(\ref{is}), as previously discussed using Bloch sphere visualization.

\section{Ideal Dispersive Case}\label{idc}
For the dispersive protocol, we define the detuning as $\delta = \omega - (E_s - E_r) = \omega - (E_{\tilde e} - E_{\tilde g}) \neq 0$. As in the resonant protocol, we assume $\lambda_i = \lambda_{sq} = \lambda$. However, unlike the resonant case, the Hamiltonian (\ref{H1}) in the interaction picture becomes time-dependent and takes the form
\begin{eqnarray}\label{htd}
\tilde H =\lambda\left[a^\dag e^{i\delta t} \left( |r\rangle\langle s|+|\tilde g\rangle\langle \tilde e|)+a e^{-i\delta t}(|s\rangle\langle r|+|\tilde e\rangle\langle \tilde g|\right)\right].\nonumber\\
\end{eqnarray}
In the dispersive limit $\delta \gg \lambda$, standard methods \cite{disp1,disp2,disp3} show that this Hamiltonian effectively reduces to
\begin{eqnarray} \label{disp}
H_D &=& \chi\left( |s\rangle\langle s|aa^\dag - |r\rangle\langle r|a^\dag a + |\tilde e\rangle\langle \tilde e|aa^\dag - |\tilde g\rangle\langle \tilde g|a^\dag a \right) \nonumber \\ 
&& + \chi \left( |r\rangle\langle s| \otimes |\tilde e\rangle\langle \tilde g| + |s\rangle\langle r| \otimes |\tilde g\rangle\langle \tilde e| \right),
\end{eqnarray} 
where
\begin{eqnarray}\label{chi}
\chi = \frac{\lambda^2}{\delta},
\end{eqnarray}
is the effective coupling constant between the atomic Rydberg levels and the superconducting qubit, mediated by the resonator mode. This constant also corresponds to the Stark shift in the energy levels due to the off-resonant interaction with the resonator.

Let us consider the initial state 
\begin{eqnarray}\label{psi0d}
 |\psi(0)\rangle &=& |r\rangle \otimes |\varphi\rangle_c \otimes \left[ \cos(\theta/2) |\tilde e\rangle + e^{i\phi} \sin(\theta/2) |\tilde g\rangle \right] \nonumber \\ 
&=& \cos(\theta/2) |r, \varphi, \tilde e\rangle + e^{i\phi} \sin(\theta/2) |r, \varphi, \tilde g\rangle, 
\end{eqnarray} 
which differs from Eq.~(\ref{is}) only in the resonator part, now assumed to be in a general pure state $|\varphi\rangle_c$. In the Fock basis, we write $|\varphi\rangle_c = \sum_{n} a_n |n\rangle$, with coefficients $a_n$ satisfying $\sum_n |a_n|^2 = 1$.  As we will see, the dispersive protocol works for any state $|\varphi\rangle_c$ of fixed parity, which is a distinctive feature compared to the resonant case which demands the resonator to be in the vacuum state. 

According to the Hamiltonian in Eq.(\ref{disp}), the initial state in Eq.(\ref{psi0d}) evolves into
\begin{eqnarray}\label{sdisp}
|\psi(t)\rangle&=&\sum_n [c_{\tilde en}(t)|r,n,\tilde e\rangle+c_{\tilde gn}(t)|r,n,\tilde g\rangle\nonumber\\ 
&&+c_{sn}(t)|s,n, \tilde g\rangle],\nonumber\\
\end{eqnarray}  
with
\begin{eqnarray}
c_{\tilde en}(t)&=&\frac{1}{2}a_n(e^{-2i\chi t}+1)\cos(\theta/2),\nonumber\\
c_{\tilde gn}(t)&=&a_n\,e^{2in\chi t}\,e^{i\phi}\sin(\theta/2) ,\nonumber\\
c_{sn}(t)&=&\frac{1}{2}a_n(e^{-2i\chi t}-1)\cos(\theta/2).
\end{eqnarray}
This allows us to determine for which states $|\varphi\rangle_c$ the protocol works. The time dependence of the component $c_{gn}(t)$ varies with the number of photons, which generally prevents the resonator field state from factorizing. However, it is still possible to identify families of field states for which this factorization is possible. Specifically, at time $t = \tau_D$, where $\chi \tau_D = \pi/2$, the coefficients become $c_{\tilde en}(\tau_D) = 0$, $c_{\tilde gn}(\tau_D) = a_n e^{in\pi} e^{i\phi}\sin(\theta/2)$, and $c_{rn}(\tau_D) = -a_n\cos(\theta/2)$, leading to
\begin{widetext}
\begin{eqnarray}
|\psi\rangle_{\tau_D} &=& \sum_n [ a_n\,e^{in\pi}e^{i\phi}\sin(\theta/2) |r,n,\tilde g\rangle  - a_n\cos(\theta/2) |s,n, \tilde g\rangle ]\nonumber\\
&=& [e^{i\phi}\sin(\theta/2)|r\rangle - \cos(\theta/2)|s\rangle] \otimes  \sum_{n\,\text{even}} a_n  |n\rangle  \otimes |\tilde g\rangle  - \left[e^{i\phi}\sin(\theta/2)|r\rangle + \cos(\theta/2)|s\rangle\right] \otimes  \sum_{n\,\text{odd}} a_n  |n\rangle  \otimes |\tilde g\rangle\nonumber\\
\end{eqnarray}
\end{widetext}
Interestingly, for resonator states where $a_n = 0$ for odd $n$ or $a_n = 0$ for even $n$, the state can be factorized. For even states, $a_n = 0$ for odd $n$, such as the well-known squeezed vacuum state \cite{disp2}, we find
\begin{eqnarray}\label{psi3} 
|\psi\rangle_{\tau_D} = \left[e^{i\phi}\sin(\theta/2)|r\rangle - \cos(\theta/2)|s\rangle\right] \otimes |\varphi\rangle \otimes|\tilde g\rangle. 
\end{eqnarray} 
Just as in the resonant case, the lasers perform the rotations $|s\rangle \to -i|e\rangle$ and $|r\rangle \to i|g\rangle$, achieved by setting $\Omega T_D = \pi/2$ and $\tilde\Omega T_D = 3\pi/2$, respectively, which once again leads to the condition $\tilde\Omega = 3\Omega$. The final state after this second step of the protocol is
\begin{eqnarray}\label{f2}
|\psi\rangle_D=\left[\cos(\theta/2)|e\rangle+ie^{i\phi}\sin(\theta/2)|g\rangle\right]\otimes|\varphi\rangle\otimes|\tilde g\rangle,
\end{eqnarray}
which represents perfect transduction. 

To investigate how the detuning $\delta$ influences the fidelity of the protocol, we numerically solved the von Neumann equation in the time interval $0 \leq t \leq \tau_D + T_D$, using the time-dependent Hamiltonian
\begin{eqnarray} \label{ht2}
H(t) &=& \tilde H + u(t-\tau_{D})\Omega (|e\rangle \langle s| + |s\rangle \langle e|) \nonumber \\ && + u(t-\tau_{D})\tilde\Omega (|r\rangle \langle g| + |g\rangle \langle r|),
\end{eqnarray}
where $\tilde H$ is the interaction-picture Hamiltonian from Eq.~(\ref{htd}), valid for arbitrary detuning $\delta$, and $u(x)$ denotes the Heaviside step function.

\begin{figure}[t!]
    \includegraphics[width=\linewidth]{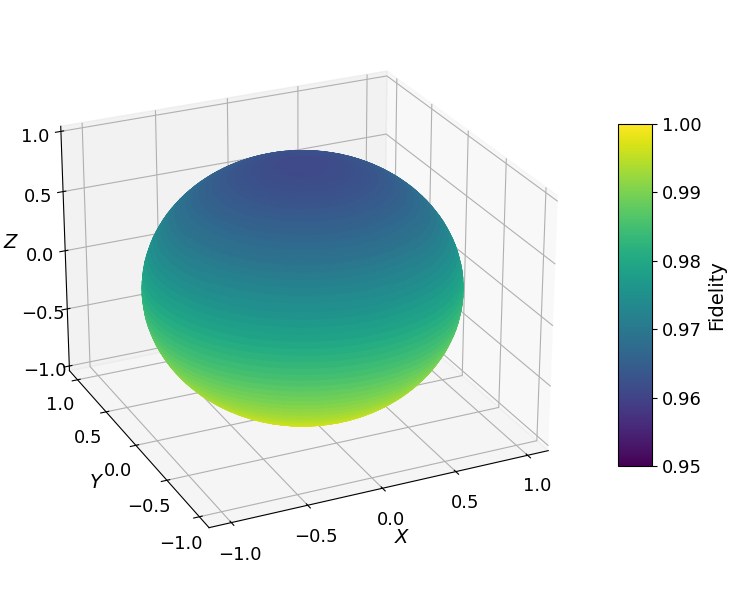}
    \includegraphics[width=\linewidth]{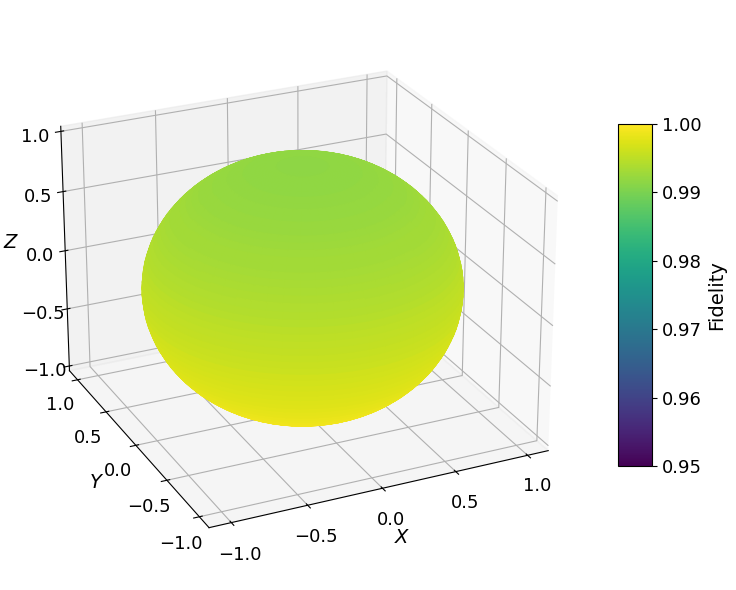}
     \caption{Bloch sphere visualization of the fidelity of the dispersive protocol at $t = \tau_D + T_D$ for $\lambda = 2\pi \times 8$ MHz and $\Omega / 2\pi = 4.8$ MHz ($\Omega / \lambda = 0.6$), with $\delta = 6\lambda$ in the top panel and $\delta = 12\lambda$ in the bottom panel.}	\label{bloch_d}
\end{figure}

In Fig.~\ref{bloch_d}, we illustrate how the fidelity depends on the detuning $\delta$, with fixed values $\lambda = 2\pi \times 8$ MHz and $\Omega = 2\pi \times 4.8$ MHz ($\Omega / \lambda = 0.6$).
In the upper panel of Fig. \ref{bloch_d}, we show the results for $\delta = 6\lambda$, which does not satisfy $\delta \gg \lambda$, but still achieves a minimum fidelity of $F_{\text{min}} = 0.96$.
When the detuning is increased to $12\lambda$ (bottom panel), the system operates in the dispersive regime, achieving $F_{\text{min}} = 0.99$. 
One remark is worth making at this point. As we have just seen, the protocol works well when $\Omega < \lambda$. This is explained by the fact that, in the dispersive regime, the relevant comparison is between $\Omega$ and the effective coupling constant $\chi$ given by Eq.(\ref{chi}). For example, when considering $\delta = 12\lambda$ and the parameters used in Fig.\ref{bloch_d}, one finds $\Omega/\chi = 7.2$. In other words, the role of the cavity during the laser application is suppressed by a sufficiently large $\Omega$. However, contrary to the resonant case, where increasing the Rabi frequency $\Omega$ with a fixed $\lambda$ always helps, this is not necessarily true in the dispersive protocol. It would be if the dispersive Hamiltonian in Eq.(\ref{disp}) were exact. Since it is not, the use of the microscopic Hamiltonian in Eq.(\ref{htd}) in the simulations may lead to small deviations from the expected monotonicity with $\Omega$. This is a consequence of the fact that $\tau_D$, by definition, is calculated from $\chi$, which belongs to the effective model. Therefore, some non-monotonicities in the dependence of the fidelity on the Rabi frequency $\Omega$ should be expected. This behavior is indeed observed in the next section, where we vary $\Omega$ and $\lambda$ independently. Finally, in neither protocol does the regime $\Omega \ll \lambda$ perform well, as we enter a limit where the influence of the cavity cannot be neglected during the application of the laser pulses.

\begin{figure}[t!]
    \includegraphics[width=\linewidth]{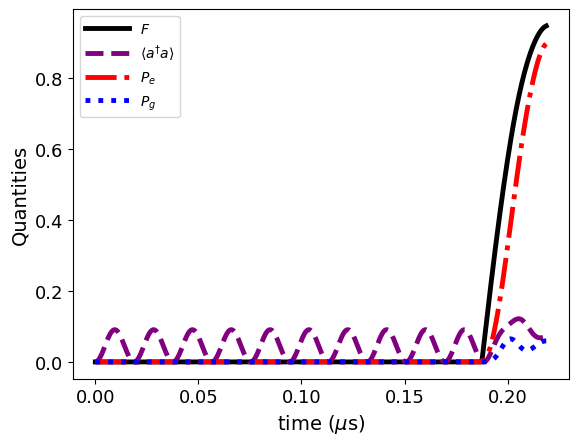}
     \includegraphics[width=\linewidth]{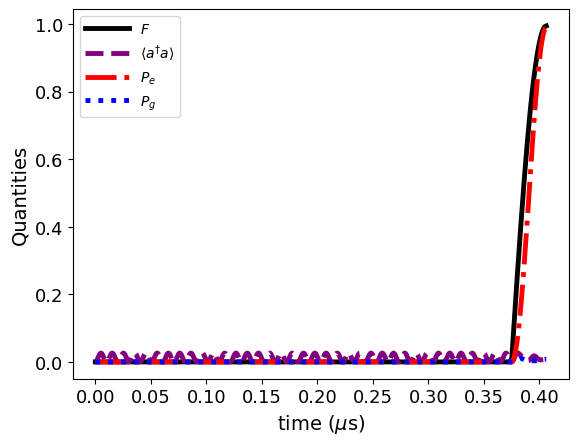}
     \caption{Dynamics of key quantities for the dispersive transduction protocol for $0\leq t \leq \tau_D + T_D$: fidelity (solid black), mean resonator photon number (dashed purple), population in $|e\rangle$ (dash-dotted red), and in $|g\rangle$ (dotted blue).  In both panels, we fix $\lambda = 2\pi \times 8$ MHz and $\Omega / 2\pi = 24$ MHz ($\Omega / \lambda = 3$).  Top panel: $\delta=6\lambda$. Bottom panel: $\delta=12\lambda$. Initial state from Eq.~(\ref{psi0d}) with $\theta = 0$.}
	\label{delta_e}
\end{figure}

To gain further insight into the dynamics of the system, we analyze the fidelity for the same two representative initial states considered in the resonant case. Specifically, we consider Eq.~(\ref{psi0d}) with $\theta = 0$ and $\theta = \pi/2$, $\phi = 0$, assuming that the resonator is initially in the vacuum state $|\phi\rangle_c = |0\rangle_c$ in both cases. The result for $\theta = 0$ is presented in Fig.~\ref{delta_e}. As shown in the upper panel, with $\delta = 6 \lambda$ a transfer fidelity of approximately 0.95 is achieved. This deviation from unity arises because the actual evolving state is not fully described by Eq.(\ref{sdisp}). A small detuning allows some excitation of the resonator mode during the state transfer process, as evidenced by the non-zero photon number $\langle a^\dagger a \rangle$. Consequently, the populations in states $|e\rangle$ and $|g\rangle$ are also affected. In particular, the population in $|e\rangle$ is expected to reach unity at the end of the protocol, but this does not occur due to residual excitation of the resonator mode. In the bottom panel of Fig.~\ref{delta_e}, the detuning is increased to $\delta = 12 \lambda$, which significantly improves the protocol's performance. The excitation of the resonator is then strongly suppressed and remains effectively in the vacuum state with $\langle a^\dag a \rangle \approx 0$ throughout the protocol. Therefore, both the population in $|e\rangle$ and the fidelity approach unity.

\begin{figure}[t!]
    \includegraphics[width=\linewidth]{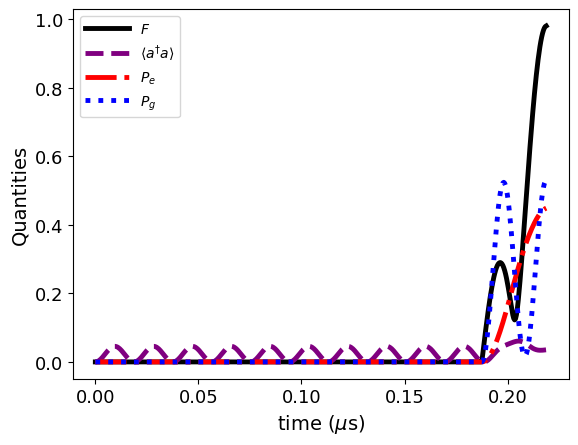}
     \includegraphics[width=\linewidth]{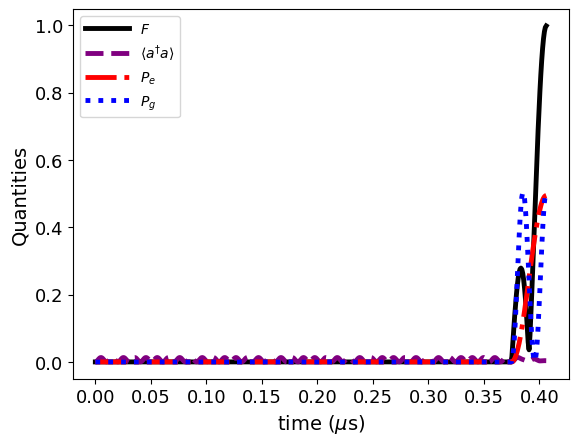}
     \caption{Dynamics of key quantities for the dispersive transduction protocol for $0\leq t \leq \tau_D + T_D$: fidelity (solid black), mean resonator photon number (dashed purple), population in $|e\rangle$ (dash-dotted red), and in $|g\rangle$ (dotted blue).  In both panels, we fix $\lambda = 2\pi \times 8$ MHz and $\Omega / 2\pi = 24$MHz ($\Omega / \lambda = 3$).  Top panel: $\delta=6\lambda$. Bottom panel: $\delta=12\lambda$. Initial state from Eq.~(\ref{psi0d}) with $\theta = \pi/2, \phi = 0$.}
	\label{delta_eg}
\end{figure}

The second case ($\theta = \pi/2$, $\phi = 0$) is shown in Fig.~\ref{delta_eg}. Once again, high fidelity is achieved even for a relatively small detuning ($\delta = 6 \lambda$), as presented in the top panel. The infidelity originates from off-resonant excitation of the resonator mode during the state transfer, which is also reflected in the oscillations of the mean photon number. As a consequence, the populations in $|e\rangle$ and $|g\rangle$ are not equal at the end of the protocol, as should ideally be. As in the previous case, this imbalance is substantially reduced by increasing the detuning to $\delta = 12 \lambda$, as shown in the bottom panel of Fig.\ref{delta_eg}. The resulting dynamics closely follow those predicted by the effective Hamiltonian in Eq.~(\ref{disp}), leading to a fidelity close to unity.

\section{Non-Ideal Dynamics: Impact of dissipation and decoherence}\label{noise}
In the previous sections, we have shown that high-fidelity state transfer is possible under idealized conditions, both in the resonant and dispersive regimes. However, in a realistic implementation, imperfections such as resonator decay, finite atomic lifetimes, and dissipation and dephasing in the superconducting qubit must be taken into account. To assess the robustness of the protocols under these conditions, we now include dissipation and decoherence in our model. Specifically, we consider resonator decay at rate $\kappa$, spontaneous emission from Rydberg levels with rates $\gamma_j$ ($j = r, s$), as well as relaxation and dephasing of the superconducting qubit at rates $\gamma_{sq}$ and $\gamma_\varphi$, respectively. The system dynamics is then governed by the master equation 
\begin{eqnarray}\label{master}
\dot\rho&=&-i[H(t),\rho]+\kappa\mathcal{D}[a]\rho+\gamma_r\mathcal{D}[\sigma_r]\rho+\gamma_s\mathcal{D}[\sigma_s]\rho \nonumber\\
&&+\gamma_{sq}\mathcal{D}[\sigma_{sq}]\rho+\gamma_{\varphi}\mathcal{D}[\tilde\sigma_{z}]\rho
\end{eqnarray}
where $\sigma_r = |g\rangle\langle r|$, $\sigma_s = |e\rangle\langle s|$, $\sigma_{sq} = |\tilde g\rangle\langle\tilde e|$, and $\tilde\sigma_z = |\tilde e\rangle\langle \tilde e| - |\tilde g\rangle\langle \tilde g|$. The term
$$
\mathcal{D}[O]\rho=O\rho O^\dag-\frac{1}{2}O^\dag O\rho-\frac{1}{2}\rho O^\dag O,
$$
represents the standard Lindblad dissipator, and the time-dependent Hamiltonian $H(t)$ is given by Eq.~(\ref{ht}) for the resonant case and by Eq.~(\ref{ht2}) for the dispersive one. Although the choice $\delta = 12\lambda$ for the dispersive protocol showed excellent agreement with the effective Hamiltonian in Eq.(\ref{disp}) in the ideal case, we use the full Hamiltonian in Eq.(\ref{ht2}) for the simulations that follow. This is because the resonator can occasionally become populated, which is a feature not captured by the effective Hamiltonian in Eq.(\ref{disp}). Accounting for this is important when evaluating the impact of resonator decay, as described by the master equation in Eq.(\ref{master}).

Realistic decay rates for the resonator mode, $\kappa/2\pi$, are on the order of a few MHz \cite{kappa1, kappa2}. For example, in \cite{kappa2}, the resonator has a fundamental frequency of 5.4 GHz and a linewidth of $\kappa/2\pi = 1.0$ MHz. For Rydberg levels, the decay rates $\gamma_s$ and $\gamma_r$ are on the order of a few kilohertz, depending on the specific Rydberg level and the ion or atom species \cite{kaler, kappa1}. In superconducting qubits, various designs of ``artificial atoms''  are integrated into circuit QED setups \cite{blais, wallaceexp}, with typical values for $\gamma_{sq}$ and $\gamma_\varphi$ around tenths and hundreds of kHz, respectively. In \cite{kappa2}, for example, the values are $\gamma_{sq}/2\pi=35$ kHz and $\gamma_\varphi/2\pi=130$ kHz, respectively. We do not include the dephasing of Rydberg levels, as it is primarily induced by laser power fluctuations, which can be minimized \cite{markus}. Thus, the primary decoherence mechanisms that affect the Rydberg states are their finite lifetimes, represented by $\gamma_s$ and $\gamma_r$. Overall, the dominant source of decoherence in our system is the resonator decay, which is about one order of magnitude larger than any of the other noise sources. Hence, we represent the transfer laser Rabi frequency and the resonator coupling in terms of $\kappa$.

To evaluate how dissipation affects the fidelity of the resonant and dispersive protocols, we solve the master equation in Eq. $(\ref{master})$ using the physical parameters outlined above.  The density matrix of the system evolves up to times $t_R = \tau_R + T_R$ for the resonant case and $t_D = \tau_D + T_D$ for the dispersive case. The fidelity at these final times is computed using Eq. $(\ref{F})$, varying independently both the coupling constant $\lambda $ and the laser Rabi frequency $\Omega$.

\begin{figure}[]
    \includegraphics[width=0.9\linewidth]{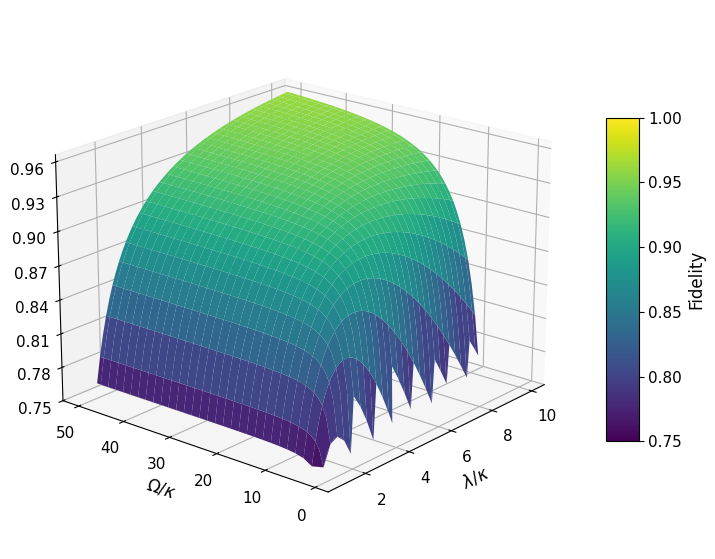}
    \includegraphics[width=0.9\linewidth]{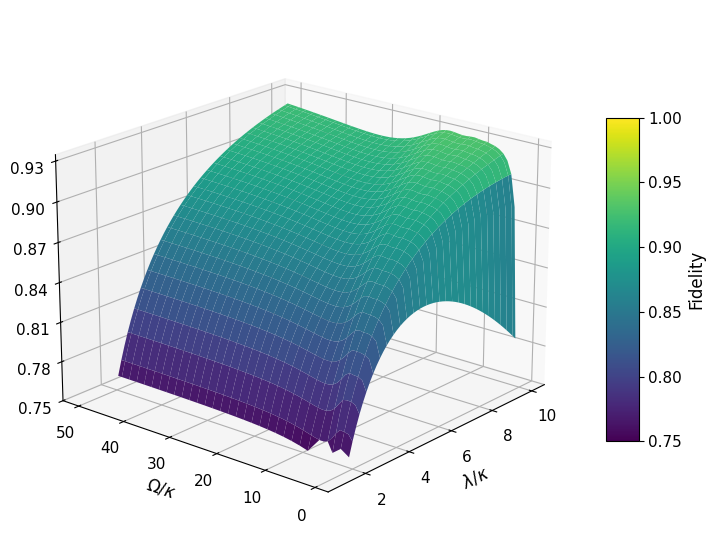}
     \caption{Fidelity as a function of the resonator coupling strength $\lambda$ and the laser Rabi frequency $\Omega$ for the noisy case with $\kappa/2\pi = 1.0$ MHz, $\gamma_s/2\pi = \gamma_r/2\pi = 1.0$ kHz, $\gamma_{sq}/2\pi = 35$ kHz, and $\gamma_\varphi/2\pi = 130$ kHz. The initial state is prepared with $\theta = 0$ in Eq.(\ref{is}) for the resonant protocol (upper panel) and in Eq.(\ref{psi0d}) for the dispersive protocol (bottom panel), where $|\varphi\rangle_c = |0\rangle_c$. The detuning for the dispersive case is fixed at $\delta = 12\lambda$.}
	\label{noise_e}
\end{figure}

\begin{figure}[]
     \includegraphics[width=0.9\linewidth]{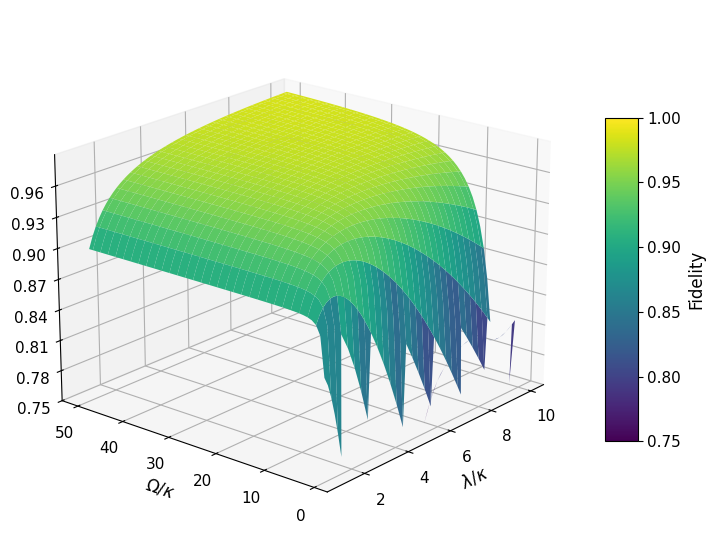}
    \includegraphics[width=0.9\linewidth]{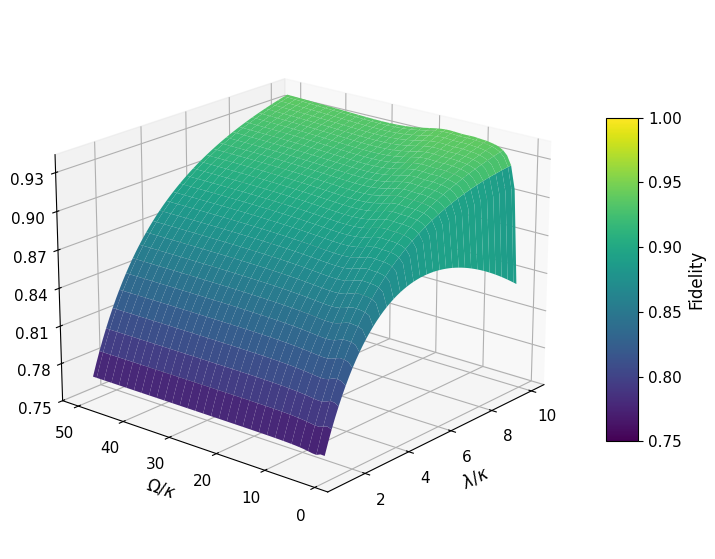}
     \caption{Fidelity as a function of the resonator coupling strength $\lambda$ and the laser Rabi frequency $\Omega$ for the noisy case with $\kappa/2\pi = 1.0$ MHz, $\gamma_s/2\pi = \gamma_r/2\pi = 1.0$ kHz, $\gamma_{sq}/2\pi = 35$ kHz, and $\gamma_\varphi/2\pi = 130$ kHz. The initial state is prepared with $\theta = \pi/2$, $\phi = 0$ in Eq.(\ref{is}) for the resonant protocol (upper panel) and in Eq.(\ref{psi0d}) for the dispersive protocol (bottom panel), where $|\varphi\rangle_c = |0\rangle_c$. The detuning for the dispersive case is fixed at $\delta = 12\lambda$.}
	\label{noise_eg}
\end{figure}

Building on the previous analysis, we consider two illustrative cases given by Eqs.~(\ref{is}) and (\ref{psi0d}): $\theta = 0$, shown in Fig.~\ref{noise_e}, and $\theta = \pi/2$, $\phi = 0$, shown in Fig.~\ref{noise_eg}. In both figures, the upper panels correspond to the resonant protocol, while the lower panels show the dispersive one. For clarity, we display only the regions where the fidelity exceeds 0.75. A common feature across all panels is that small values of $\Omega$ and $\lambda$ degrade the fidelity of the protocol. Naturally, in these regimes, cavity decay dominates the dynamics. However, another important source of infidelity emerges when $\Omega \ll \lambda$: the resonator can become repopulated with photons during the atomic state transfer, which reduces the overall fidelity. This effect has been discussed in previous sections.

For the dispersive regime with both initial states, a nonmonotonic dependence of the fidelity on $\Omega$, for fixed $\lambda$, is observed when $\Omega$ is comparable to $\lambda$. This is more evident for the preparation considered in Fig.~\ref{noise_e}, where a particularly noticeable plateau appears around $\lambda/\kappa \approx 10$. As explained previously, this behavior results from the small discrepancy between the effective Hamiltonian in Eq.~(\ref{disp}), used to determine $\tau_D$, and the full Hamiltonian in Eq.~(\ref{ht}), used in the simulations. As also seen in Fig.~\ref{noise_e}, increasing $\Omega$ while keeping $\lambda$ fixed gradually restores the expected monotonic behavior.

From a practical perspective, it is noteworthy that, under realistic dissipation and decoherence, a wide range of feasible values for $\Omega$ and $\lambda$ yield fidelities above 0.9 for both protocols.
This shows that high-fidelity state transfer does not require extreme coupling strengths. In Fig.~\ref{noise_e}, for example, with $\Omega/\kappa \approx 5$ as reported in Ref.\cite{Lahaye}, we find $F_{\text{max}} = 0.85$ at $\lambda/\kappa \approx 3$ for the resonant protocol, while the dispersive case yields $F_{\text{max}} = 0.93$ at $\lambda/\kappa \approx 10$. Similarly, Fig.~\ref{noise_eg} shows $F_{\text{max}} = 0.94$ at $\lambda/\kappa \approx 2.8$ for the resonant case and at $\lambda/\kappa \approx 10$ for the dispersive case. Although the resonant protocol generally yields slightly higher fidelities when $\Omega$ and $\lambda$ are not constrained, the dispersive protocol tends to perform better at lower Rabi frequencies. This behavior is again related to the fact that, in the dispersive regime, the role of the cavity during the laser application depends on the relationship between $\Omega$ and the effective coupling constant $\chi$, as given by Eq.~(\ref{chi}) and discussed in Sect.~\ref{idc}.

\begin{figure}[t!]
    \centering
    \includegraphics[width=0.9\linewidth]{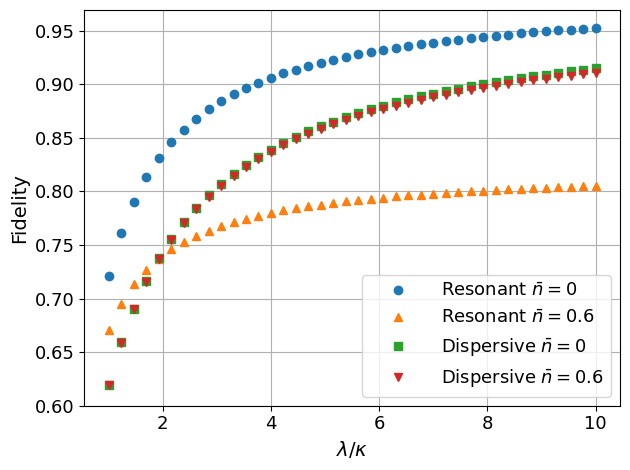}
    \includegraphics[width=0.9\linewidth]{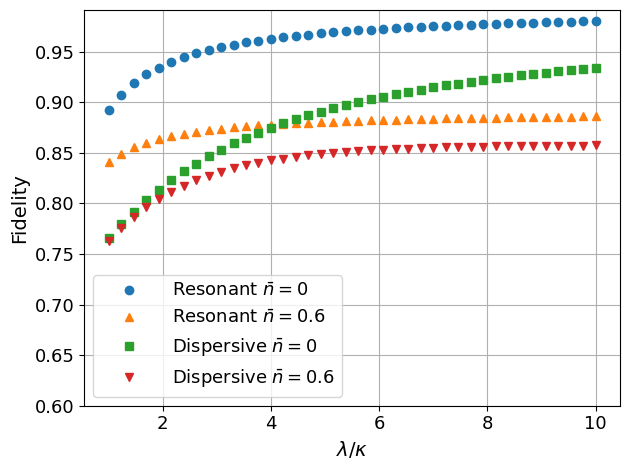}
    \caption{Fidelity as a function of the resonator coupling strength $\lambda$ for initial thermal states of the resonator, in the presence of noise. The superconducting qubit is initialized with $\theta = 0$ in Eq. (\ref{sqstate}) for the top panel, and with $\theta = \pi/2$, $\phi = 0$ in Eq. (\ref{sqstate}) for the bottom panel. The trapped atom is prepared in the state $|r\rangle$, and the cavity in the thermal state given by Eq. (\ref{ther}), with $\bar{n} = 0$ and $\bar{n} = 0.6$ (see legend). The remaining parameters are $\kappa/2\pi = 1.0$ MHz, $\gamma_s/2\pi = \gamma_r/2\pi = 1.0$ kHz, $\gamma_{sq}/2\pi = 35$ kHz, $\gamma_\varphi/2\pi = 130$ kHz, and $\Omega = 3\lambda$. For the dispersive case, the detuning is fixed at $\delta = 12\lambda$.}
    \label{thermal}
\end{figure}

In superconducting systems thermal excitation may not be negligible. Typically, instead of vacuum, the resonator is in a thermal state given by
\begin{equation}\label{ther}
\rho_{\text{th}} = \sum_{n=0}^\infty \frac{\bar{n}^n}{(\bar{n}+1)^{n+1}} \, |n\rangle \langle n|,
\end{equation}
where $\bar{n}$ is the mean thermal photon number. In state-of-the-art experiments, the resonator temperatures are low enough to guarantee $\bar{n} \ll 1$, so that typically only a very small residual excitation is present in the resonator (e.g. $\bar{n} = 0.006$ as reported in \cite{nt}). At this level, thermal occupation has a negligible impact on the performance of our protocols. However, to better illustrate the role of the thermal resonator population, we consider an extreme scenario with $\bar{n} = 0.6$, which is two orders of magnitude higher than in \cite{nt}. We examine the scenarios in Fig.~\ref{thermal}, this time with $\Omega = 3\lambda$. Also, for the dispersive case, we fix $\delta = 12\lambda$. As before, the atomic system is initially prepared in the Rydberg state $|r\rangle$, but now the resonator is in the thermal equilibrium state in Eq.~(\ref{ther}). 

For the superconducting qubit initially in the $\theta = 0$ state [see Eq.(\ref{sqstate})], the resonant protocol (upper panel in Fig.\ref{thermal}) shows a significant drop in fidelity due to thermal photons, whereas the dispersive protocol exhibits almost no degradation. The situation changes for the superposition state with $\theta = \pi/2$ and $\phi = 0$: in both protocols, the fidelity is significantly reduced by the thermal population of the resonator. However, there is one important difference: In the dispersive case, the effect of thermal photons becomes more pronounced for stronger cavity couplings. Further simulations have shown that the strong dependence on the initial state in the dispersive regime is not the result of a single physical factor, such as the sole presence of thermal photons. Instead, the inability to factorize the photonic state [see Eq.(\ref{psi3})], caused by thermal photons, together with the various dissipation and decoherence mechanisms considered in our model, leads to effective dephasing dependent on $\lambda$ that strongly impacts the superposition state, as seen in Fig.\ref{thermal}. A further layer of complexity in understanding this behavior lies in the fact that the dynamics in the dispersive protocol do not correspond to a perfect dispersive interaction as given by Eq.(\ref{disp}). Small oscillations in the cavity population, combined with dephasing and dissipation, contribute to the effective $\lambda$-dependent dephasing affecting the fidelity as observed in Fig.\ref{thermal}.

\section{Conclusions}\label{sumup}
In this work, we have presented two protocols, resonant and dispersive, to transfer quantum states from superconducting qubits to electronic levels of trapped atoms or ions, mediated by a microwave resonator. These protocols utilize excited Rydberg levels of the trapped atomic particle, which interact indirectly with superconducting qubits through their mutual coupling to a microwave resonator. Laser pulses are applied to de-excite the atomic system from the Rydberg states to low-lying levels, completing the transduction.

Our work shows that both the resonant and dispersive protocols achieve high-fidelity state transfer under realistic dissipation and decoherence, with fidelities exceeding 0.9 across broad regions of the parameter space spanned by the transfer laser Rabi frequencies and the resonator coupling strength. Although the resonant protocol generally performs slightly better in ideal scenarios or at high Rabi frequencies, the dispersive protocol stands out for its ability to achieve high fidelities even at moderate values of $\Omega$. This makes the dispersive scheme particularly attractive in experimental settings where the Rydberg excitation Rabi frequency may be limited.

We have also examined the influence of thermal photons in the resonator, which may be relevant in superconducting platforms. While the resonant protocol is affected by thermal occupation regardless of the initial qubit state, the dispersive protocol is more sensitive to thermal photons when the superconducting qubit starts in a superposition state. This sensitivity results from a combination of thermal excitation, dissipation, and deviations from the ideal dispersive regime.

As a final remark, the ideal dispersive protocol also works with pure resonator states with defined parity, whereas the resonant case is limited to the resonator being in the vacuum state. However, this does not immediately imply an advantage, as generating such states is typically more complex than preparing the vacuum state.

\section*{ACKNOWLEDGEMENTS} 
F. L. S. acknowledges partial support from the Conselho Nacional de Desenvolvimento Científico e Tecnológico CNPq (Grant No. 313068/2023-2).
M. K. acknowledges support from the EPSRC UK Quantum Technologies Programme under grant EP/T001062/1.

\section*{DATA AVAILABILITY STATEMENT} 
Data are available on request from the authors.


\begin{thebibliography}{99}

\bibitem{PhysRevLett.102.083602}  Atac Imamo\ifmmode \breve{g}\else \u{g}\fi{}lu, Cavity QED Based on Collective Magnetic Dipole Coupling: Spin Ensembles as Hybrid Two-Level Systems, Phys. Rev. Lett. \textbf{102}, 083602 (2009).

\bibitem{PhysRevLett.103.043603} J. Verd\'u, J. H. Zoubi, C. Koller, J. Majer, H. Ritsch, and J. Schmiedmayer, Strong Magnetic Coupling of an Ultracold Gas to a Superconducting Waveguide Cavity, Phys. Rev. Lett. \textbf{103}, 043603 (2009).

\bibitem{PhysRevA.85.020302} M. Hafezi, Z. Kim, S. L. Rolston, L. A. Orozco, B. L. Lev, and J. M. Taylor, Atomic interface between microwave and optical photons, Phys. Rev. A \textbf{85}, 020302 (2012).

\bibitem{PhysRevLett.113.063603} C. O’Brien, N. Lauk, S. Blum, G. Morigi, and M. Fleischhauer, Interfacing Superconducting Qubits and Telecom Photons via a Rare-Earth-Doped Crystal, Phys. Rev. Lett. \textbf{113}, 063603 (2014).

\bibitem{PhysRevLett.113.203601} L. A. Williamson, Y. -H. Chen, and J. J. Longdell, Magneto-Optic Modulator with Unit Quantum Efficiency, Phys. Rev. Lett. \textbf{113}, 203601 (2014).

\bibitem{PhysRevA.92.062313} X. Fernandez-Gonzalvo, Y. -H. Chen, C. Yin, S. Rogge, and J. J. Longdell, Coherent frequency up-conversion of microwaves to the optical telecommunications band in an Er:YSO crystal, Phys. Rev. A \textbf{92}, 062313 (2015).

\bibitem{PhysRevLett.105.220501} K. Stannigel, P. Rabl, A. S. S\o{}rensen, P. Zoller, and M. D. Lukin, Optomechanical Transducers for Long-Distance Quantum Communication, Phys. Rev. Lett. \textbf{105}, 220501 (2010).

\bibitem{Andrews2014-bo} R. W. Andrews, R. W. Peterson, T. P. Purdy, K. Cicak, R. W. Simmonds, C. A. Regal, and K. W. Lehnert, Bidirectional and efficient conversion between microwave and optical light, Nat. Phys. \textbf{10}, 321 (2014).

\bibitem{Bagci2014-it} T. Bagci, A. Simonsen, S. Schmid, L. G. Villanueva, E. Zeuthen, J. Appel, J. M. Taylor, A. S\o{}rensen, K. Usami, A. Schliesser, and E. S. Polzik, Optical detection of radio waves through a nanomechanical transducer, Nature \textbf{507}, 81 (2014).

\bibitem{PhysRevLett.109.130503} Sh. Barzanjeh, M. Abdi, G. J. Milburn, P. Tombesi, and D. Vitali, Reversible Optical-to-Microwave Quantum Interface, Phys. Rev. Lett. \textbf{109}, 130503 (2012).

\bibitem{Safavi-Naeini_2011}  A. H. Safavi-Naeini and O. Painter, Proposal for an optomechanical traveling wave phonon–photon translator, New J. Phys. \textbf{13} 013017 (2011).

\bibitem{PhysRevA.82.053806} L. Tian and H. Wang, Optical wavelength conversion of quantum states with optomechanics, Phys. Rev. A \textbf{82}, 053806 (2010).

\bibitem{PhysRevLett.108.153603} Y. -D. Wang and A. A. Clerk, Using Interference for High Fidelity Quantum State Transfer in Optomechanics, Phys. Rev. Lett. \textbf{108}, 153603 (2012).

\bibitem{PhysRevLett.108.153604} L. Tian, Adiabatic State Conversion and Pulse Transmission in Optomechanical Systems, Phys. Rev. Lett. \textbf{108}, 153604 (2012).

\bibitem{Hill2012-y} J. T. Hill, A. H. Safavi-Naeini, J. Chan, and O. Painter, Coherent optical wavelength conversion via cavity optomechanics, Nature Communications \textbf{3}, 1196 (2012).

\bibitem{PhysRevA.81.063837} M. I. Tsang, Cavity quantum electro-optics, Phys. Rev. A \textbf{81}, 063837 (2010).

\bibitem{PhysRevA.96.043808} M. Soltani, M. Zhang, Efficient quantum microwave-to-optical conversion using electro-optic nanophotonic coupled resonators, C. A. Ryan, G. J. Ribeill, C. Wang, and M. Loncar, Phys. Rev. A \textbf{96}, 043808 (2017).

\bibitem{PhysRevB.93.174427} R. Hisatomi, A. Osada, Y. Tabuchi, T. Ishikawa, A. Noguchi, R. Yamazaki, K. Usami, and Y. Nakamura, Bidirectional conversion between microwave and light via ferromagnetic magnons, Phys. Rev. B \textbf{93}, 174427 (2016).

\bibitem{PhysRevA.99.063830} J. R. Everts, M. C. Berrington, R. L. Ahlefeldt, and J. J. Longdell, Microwave to optical photon conversion via fully concentrated rare-earth-ion crystals, Phys. Rev. A \textbf{99}, 063830 (2019).

\bibitem{PhysRevB.101.214414}  J. R. Everts, G. G. G. King, N. Lambert, S. Kocsis, S. Rogge, and J. J. Longdell, Ultrastrong coupling between a microwave resonator and antiferromagnetic resonances of rare-earth ion spins, Phys. Rev. B \textbf{101}, 214414 (2020).

\bibitem{PhysRevLett.118.140501}  S. Das, V. E. Elfving, S. Faez, and A. S. S\o{}rensen, Interfacing Superconducting Qubits and Single Optical Photons Using Molecules in Waveguides, Phys. Rev. Lett. \textbf{118}, 140501 (2017).

\bibitem{PhysRevA.100.053843} V. E. Elfving, S. Das, and A. S. S\o{}rensen, Enhancing quantum transduction via long-range waveguide-mediated interactions between quantum emitters, Phys. Rev. A \textbf{100}, 053843 (2019).

\bibitem{PhysRevB.96.165312} Y. Tsuchimoto, P. Knüppel, A. Delteil, Z. Sun, M. Kroner, and A. Imamo\ifmmode \breve{g}\else \u{g}\fi{}lu, Proposal for a quantum interface between photonic and superconducting qubits, Phys. Rev. B \textbf{96}, 165312 (2017).

\bibitem{Hensinger} D. De Motte, A. R. Grounds, M. Reh\'ak, A. Rodriguez Blanco, B. Lekitsch, G. S. Giri, P. Neilinger, G. Oelsner, E. Il'ichev, M. Grajcar, and W. K. Hensinger, Experimental system design for the integration of trapped-ion and superconducting qubit systems, Quantum Inf. Process \textbf{15}, 5385, (2016). 

\bibitem{milburn} D. Kielpinski, D. Kafri, M. J. Woolley, G. J. Milburn, and J. M. Taylor, Quantum Interface between an Electrical Circuit and a Single Atom, Phys. Rev. Lett. \textbf{108}, 130504 (2012).

\bibitem{molecular} D. I. Schuster, Lev S. Bishop, I. L. Chuang, D. DeMille, and R. J. Schoelkopf, Cavity QED in a molecular ion trap, Phys. Rev. A \textbf{83}, 012311 (2011).

\bibitem{Keller} T. Walker, K. Miyanishi, R. Ikuta, H. Takahashi, S. V. Kashanian, Y. Tsujimoto, K. Hayasaka, T. Yamamoto, N. Imoto, and M. Keller, Long-Distance Single Photon Transmission from a Trapped Ion via Quantum Frequency Conversion, Phys. Rev. Lett. \textbf{120}, 203601 (2018).

\bibitem{Krutyanskiy} V. Krutyanskiy, M. Canteri, M. Meraner, J. Bate, V. Krcmarsky, J. Schupp, N. Sangouard e B. P. Lanyon,  Telecom-Wavelength Quantum Repeater Node Based on a Trapped-Ion Processor, Phys. Rev. Lett. \textbf{130}, 213601 (2023).

\bibitem{haroche} J. M. Raimond, M. Brune, and S. Haroche, Manipulating quantum entanglement with atoms and photons in a cavity, Rev. Mod. Phys. \textbf{73}, 565 (2001).

\bibitem{wilk} T. Wilk, A. Ga\"{e}tan, C. Evellin, J. Wolters, Y. Miroshnychenko, P. Grangier, and A. Browaeys, Entanglement of Two Individual Neutral Atoms Using Rydberg Blockade, Phys. Rev. Lett. \textbf{104}, 010502 (2010).

\bibitem{Saffman} L. Isenhower, E. Urban, X. L. Zhang, A. T. Gill, T. Henage, T. A. Johnson, T. G. Walker, and M. Saffman, Demonstration of a Neutral Atom Controlled-NOT Quantum Gate, Phys. Rev. Lett. \textbf{104}, 010503 (2010).

\bibitem{kaler} T. Feldker, P. Bachor, M. Stappel, D. Kolbe, R. Gerritsma, J. Walz, and F. Schmidt-Kaler, Rydberg Excitation of a Single Trapped Ion, Phys. Rev. Lett. \textbf{115}, 173001 (2015).

\bibitem{markusprl} G. Higgins, F, Pokorny, C. Zhang, Q. Bodart, and M. Hennrich, Coherent Control of a Single Trapped Rydberg Ion, Phys. Rev. Lett. \textbf{119}, 220501 (2017).

\bibitem{markus} C. Zhang, F. Pokorny, W. Li, G. Higgins, A. P\"{o}schl, I. Lesanovsky, and M. Hennrich, Submicrosecond entangling gate between trapped ions via Rydberg interaction, Nature \textbf{580}, 345-349 (2020).

\bibitem{sillanpaa} M. A. Sillanp\"{a}\"{a}, J. I. Park, and R. W. Simmonds, Coherent quantum state storage and transfer between two phase qubits via a resonant cavity, Nature \textbf{449}, 438 (2007).

\bibitem{blais} A. Blais, A. Grimsmo, S. M. Girvin, and A. Wallraff, Circuit quantum electrodynamics, Rev. Mod. Phys. \textbf{93}, 025005 (2021).

\bibitem{Kaiser} M. Kaiser, C. Glaser, L Y. Ley, J. Grimmel, H. Hattermann, D. Bothner, D. K\"{o}lle, R. Kleiner, D. Petrosyan, A. Günther, and J. Fort\'agh, Cavity-driven Rabi oscillations between Rydberg states of atoms trapped on a superconducting atom chip, Phys. Rev. Research \textbf{4}, 013207 (2022).

\bibitem{NC} M. A. Chuang and Isaac L. Chuang, \textit{Quantum Computation and Quantum Information}, (Cambridge University Press, Cambridge, 2000).

\bibitem{Lahaye}  S. de Lsleuc, D. Barredo, V. Lienhard, A. Browaeys, and T. Lahaye, Analysis of imperfections in the coherent optical excitation of single atoms to Rydberg states, Phys. Rev. A \textbf{97}, 053803 (2018).

\bibitem{kappa2} A. M. Gunyhó, S. Kundu, J. Ma, W. Liu, S. Niemelä, G. Catto, V. Vadimov, V. Vesterinen, P. Singh, Q. Chen, and M. Möttönen, Single-shot readout of a superconducting qubit using a thermal detector, Nature Electronics \textbf{7}, 288 (2024). 

\bibitem{disp1} S. B. Zheng and G. C. Guo, Efficient Scheme for Two-Atom Entanglement and Quantum Information Processing in Cavity QED, Phys. Rev. Lett. \textbf{85}, 2392 (2000).

\bibitem{disp2} C. Gerry and P. Knight, \textit{Introductory Quantum Optics}, (Cambridge University Press, Cambridge, 2005).

\bibitem{disp3} F. L. Semi\~ao, Single-mode two-channel cavity QED, J. Phys. B: At. Mol. Opt. Phys. \textbf{41}, 081004  (2008).

\bibitem{kappa1} R. J. Schoelkopf and S. M. Girvin, Wiring up quantum systems, Nature \textbf{451}, 664 (2008).

\bibitem{wallaceexp}  W. Teixeira, T. M\"{o}rstedt, A. Viitanen,  H. Kivij\"{a}rvi, A. Gunyh\'o, M. Tiiri, S. Kundu, A. Sah, V. Vadimov, and M. M\"{o}tt\"{o}nen, Many-excitation removal of a transmon qubit using a single-junction quantum-circuit refrigerator and a two-tone microwave drive, Sci. Rep. \textbf{14}, 13755 (2024).

\bibitem{nt} Yan, F., Gustavsson, S., Kamal, A. et al. The flux qubit revisited to enhance coherence and reproducibility. Nat Commun \textbf{7}, 12964 (2016).

\bibitem{Chew} Y. Chew, T. Tomita, T.P.Mahesh, S. Sugawa, S. de L{\'e}s{\'e}leuc, and K. Ohmori, K., Ultrafast energy exchange between two single Rydberg atoms on a nanosecond timescale, Nature Photonics \textbf{16}, 724 (2022).

\bibitem{Huber} B. Huber, T. Baluktsian, M. Schlagm\"uller, A. K\"olle, H. K\"ubler, R. L\"ow, and T. and Pfau, GHz Rabi Flopping to Rydberg States in Hot Atomic Vapor Cells,
Phys. Rev. Lett. \textbf{107}, 243001 (2011).

\end{thebibliography}
\end{document}